\documentclass[twocolumn,showpacs]{revtex4}

\usepackage{amssymb}
\usepackage[dvips]{graphicx}

\begin{document}

\title{Unconventional high-temperature  superconductivity from repulsive interactions: theoretical constraints}
\author{A. S. Alexandrov and V. V. Kabanov}
\affiliation{Department of Physics, Loughborough University,
Loughborough LE11 3TU, United Kingdom and Josef Stefan Institute,
1001 Ljubljana, Slovenia \\}

\begin{abstract}
Unconventional   symmetries of the order
 parameter allowed  some researchers to maintain that a purely repulsive
interaction between electrons  provides superconductivity without
phonons in a number of  high-temperature superconductors. It is
shown that the Cooper pairing in p and d states is not possible with
the realistic Coulomb repulsion between fermions at  relevant
temperatures in any dimension.
\end{abstract}

\pacs{71.38.-k, 72.15.Jf, 74.72.-h, 74.25.Fy}

\maketitle

In  the theoretical analysis, the pairing mechanism of carriers
 could be not only phononic as in the BCS theory  or its
strong-coupling bipolaronic extension \cite{alebook}, but also
excitonic, plasmonic,
 magnetic, kinetic, or due to some purely repulsive
 interaction combined with the unconventional pairing symmetry of the order
 parameter.
Actually, following the original proposal by P. W. Anderson, many
authors \cite{tJ} assumed that the electron-electron interaction in
high-temperature superconductors was  strong but repulsive providing
high $T_{c}$ without  phonons via  superexchange and/or
spin-fluctuations in the d-wave pairing channel ($l=2$). A
motivation for this concept can be found in the earlier work by Kohn
and Luttinger (KL) \cite{kohn}, who showed that the Cooper pairing
of fermions with any weak  repulsion was possible since the
two-particle interaction induced by many-body effects is attractive
for pairs with  large orbital momenta, $l\gg 1$.

While the KL work did not provide the specification of the  actual
angular momentum of condensed Cooper pairs,  Fay and Layzer
\cite{fay} found that a system of \emph{hard-sphere} fermions
condenses at low densities into a p-orbital state ($l=1$). The
critical transition temperature  T$_{c}$ of repulsive fermions was
estimated well below 0.1 K with very little enhancement due to a
flatness of the Fermi surface \cite{lut}.  In two dimensions (2D)
the KL effect is absent for the parabolic band dispersion, but the
d-wave pairing appears with the repulsive Hubbard U potential when
tight-binding corrections to the electron energy spectrum are
taken into account \cite{bar}. A few subsequent studies
 argued that the KL pairing with moderate
values of angular momenta ($p$ or $d$) was impossible for
\emph{charged} fermions with the \emph{realistic finite-range}
Coulomb repulsion in the dense limit, $r_s < 1$, where the
perturbation expansion is justified \cite{lut,alegol,bar}.
Nevertheless, formally extending the perturbation results to the
opposite dilute limit, $r_s\rightarrow \infty$, Ref.\cite{bar}
found the p-wave pairing with $T_c$ about 0.01 K.

More recent studies  claimed that weak repulsive interactions
combined with lattice-induced band-structure effects do result in
higher  values of T$_c$
 in a spin singlet d-wave channel near half-filling "encouragingly
similar to what is found in the cuprate high-temperature
superconductors" \cite{kivsca}. More surprisingly these studies
claim that "in the small $r_s$ limit, the Coulomb interactions are
sufficiently well screened that it may be reasonable to treat them
as weak and short-ranged" \cite{kiv}.

Here we show that the weak and intermediate Coulomb repulsions do
not result in the Cooper instability in p- or  d-wave channels in
3D and 2D with or without lattice-induced band-structure effects.
Based on some variational simulations we also argue  that the
repulsive hard-sphere (i.e.  Hubbard U) model does not account for
high-temperature superconductivity in the  strong-coupling regime
either.

\begin{figure}[tbp]
\begin{center}
\includegraphics[angle=-90,width=0.50\textwidth]{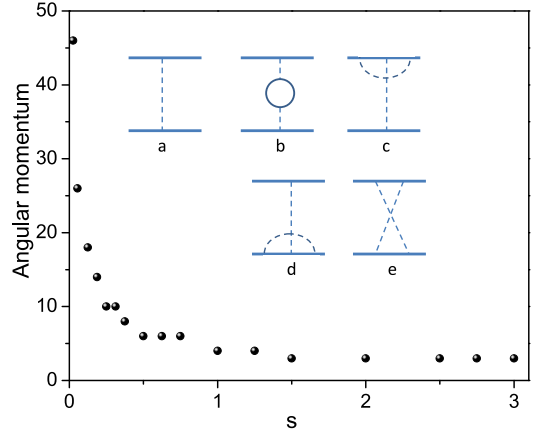}
\end{center}
\caption{(Color online) Angular momentum of the order
parameter as the function of the dimensionless repulsion $s=e^2/\pi \hbar v_F$ in the 3D Coulomb gas. Inset shows
the first order (a) and the second-order (b,c,d,e) contributions to the
two-particle vertex.} \label{one}
\end{figure}

In the framework of the BCS theory the  symmetry of the order
parameter $\Delta(\textbf{p})$ is found from the linearised integral
equation
\begin{equation}
\lambda \Delta(\textbf{p})={\Omega\over{(2\pi \hbar)^d}}\oint {dS
\over{v_F(S)}} K(\textbf{p}, \textbf{p}^\prime)
\Delta(\textbf{p}^\prime) \label{bcs}
\end{equation}
with the most negative eigenvalue $\lambda$. Here the integral is
taken over the Fermi surface (or contour in 2D), $\Omega$ is the
normalisation volume in 3D ($d=3)$ or area in 2D ($d=2$), $v_F$ is
the Fermi speed, and $K(\textbf{p}, \textbf{p}^\prime)$ is the
two-particle vertex. The sum of the first  and  second order
diagrams, Fig.(\ref{one}), for the two-particle vertex  is evaluated from
\cite{kohn,lut}:
\begin{eqnarray}
K(\textbf{p}, \textbf{p}^\prime)&=&
v(\textbf{p}-\textbf{p}^\prime)+v(\textbf{p}-\textbf{p}^\prime)
\sum_\textbf{k}[2v(\textbf{p}-\textbf{p}^\prime) \cr &-&
v(\textbf{k}+\textbf{p}^\prime)-v(\textbf{k}-\textbf{p})]Q(\textbf{p}-\textbf{p}^\prime,\textbf{k})\cr
&-&\sum_\textbf{k}v(\textbf{p}-\textbf{k})
v(\textbf{k}+\textbf{p}^\prime)Q(\textbf{p}+\textbf{p}^\prime,\textbf{k})
\label{vertex}
\end{eqnarray}
with
$Q(\textbf{q},\textbf{k})=(f_\textbf{k}-f_{\textbf{k}-\textbf{q}})/(E_\textbf{k}-E_{\textbf{k}-\textbf{q}})$,
$f_\textbf{k}$ is the Fermi-Dirac distribution function, and
$E_\textbf{k}$ is the Bloch-band dispersion.

The second-order terms in Eq.(\ref{vertex}) are prohibitively
difficult to evaluate when the Fourier transform of the repulsive
potential, $v(\textbf{q})$ depends on $q$, so that most previous
and recent studies \cite{kivsca,kiv} confined to the hard-sphere
(Hubbard U) repulsion with $v(\textbf{q})$=constant. In this case
the first order does not contribute to $\lambda$ in any
unconventional channel with $l\geqslant 1$, and the only
(attractive) contribution comes from the last sum in
Eq.(\ref{vertex}) providing p-wave pairing in 3D for parabolic
dispersion \cite{fay} and d-wave pairing in 2D for the
 tight-binding dispersion \cite{bar,kivsca,kiv}.

 Different from
those studies we consider here a more realistic Coulomb repulsion
with
 $v(\textbf{q})=4\pi e^2/\Omega (q^2+\kappa^2)$ in 3D and $v(\textbf{q})=2\pi e^2/\Omega
 (q+\kappa)$ in 2D, where $\kappa$ is the inverse screening length.
 To elucidate the role of a finite potential radius we  first take $\kappa$ as an
 independent large parameter $\kappa \gg k_F$, where $\hbar k_F$ is the Fermi momentum. Such a short screening length could be due to another component of heavy carriers.  In this
 limit, which  mimics  the Hubbard U model, one can neglect the $q$ dependence of the potential  in the second-order
 diagrams, while taking it into account  in the first-order contribution. Then the first sum in Eq.(\ref{vertex}) cancels. In the
 case of  the 3D parabolic dispersion $E_\textbf{k}=k^2/2m$ one can expand the order
 parameter  in a series of the Legendre
 polynomials, $P_l(\cos \Theta)$ on the spherical Fermi surface \cite{kohn} to
 obtain from Eq.(\ref{bcs}) the $l$-channel eigenvalue as
\begin{equation}
\lambda_l=\int_0^{\pi}d\Theta \sin(\Theta)  P_l(\cos
\Theta)\Gamma(\cos \Theta),\label{lambda}
\end{equation}
with
\begin{eqnarray}
&&\Gamma(\cos \Theta) = {s \over{2(1-\cos\Theta)+(\kappa/k_F)^2}}\cr
 &+&{s^2k_F^4\over \kappa^4}{1-\cos\Theta\over{4\sqrt{2(1+\cos \Theta)}}}\ln {\sqrt{2}+\sqrt{1+\cos\Theta}\over{\sqrt{2}-\sqrt{1+\cos\Theta}}},
\end{eqnarray}
where we define the small expansion parameter as $s=e^2/(\pi \hbar
v_F) \approx r_s/6$ ($r_s$ is the dimensionless Wigner-Seitz
radius, and $v_F=\hbar k_F/m$ is the Fermi speed).

Integrating in Eq.(\ref{lambda}) over the scattering angle yields
$\lambda_1/s=(k_F/\kappa)^4[4/3-2s (2\ln2-1)/5]$ and
$\lambda_2/s=(k_F/\kappa)^4[16(k_F/\kappa)^2/15-4s(8-11
\ln2)/105]$ for  p and d -channels, respectively, at strong
screening, $\kappa/k_F \gg 1$. For the p-wave symmetry a negative
$\lambda$ appears only   if $s
> 10/[3(2\ln 2-1)] \approx 9$ \emph{ no matter} how small the
screening length is.  This surprising result is due to a
nonvanishing repulsive contribution to the p-wave channel of the
lowest-order q-correction to the Hubbard repulsion ($\propto
q^2/\kappa^4$), which is proportional to the same power in the
screening length ($\kappa^{-4}$) as the attraction. Different from
the p-channel this  lowest-order q-correction  does not contribute
to the repulsion in the d-channel, so that $\lambda_2$ becomes
negative at some critical screening, $(\kappa/k_F)^2 > 420/[15 s
(8-11 \ln2)]\approx 75/s$. However, it is unrealistic to find the
screening length
 one  order  of magnitude smaller than the
lattice constant in real solids, so  that neither p- nor d-wave
pairing can be realised in the 3D Coulomb gas in the weak- and
intermediate-coupling regimes, where $s\lesssim 1$,  if the
screening length, $\kappa^{-1}$, is chosen to be small.

Actually the screening length in the single-component dense Coulomb
gas, where the perturbation expansion in powers of $s$ makes
sense,  is \emph{large}, $(\kappa/k_F)^2=4 s \lesssim 1$, so that
it is  unreasonable to treat  the Coulomb repulsion as weak
and at the same time as short-ranged, contrary to the
suggestion of Ref.\cite{kiv}. Following Refs.\cite{alegol,bar} we
now include all buble diagrams in the screened potential replacing
the screening momentum $\kappa$ by the static Lindhard function,
\begin{equation}
\left({\kappa\over{k_F}}\right)^2 \Longrightarrow
s(q)=4s\left[{1\over{2}}+{k_F^2-q^2/4
\over{2q k_F}}\ln {k_F+q/2\over{k_F-q/2}}\right]
\end{equation}

It is sufficient to use the Fourier component of the potential
with the long-wave screening, $s(0)=4s$ and the backward
scattering $\textbf{p}^\prime = - \textbf{p}$ with
$\textbf{p}=\textbf{q}_-/2$ where
$\textbf{q}_-=\textbf{p}-\textbf{p}^\prime $ in diagrams of
Fig.(\ref{one} c,d), and the forward scattering with
$\textbf{q}_+=\textbf{p}+\textbf{p}^\prime \approx 2 \textbf{p}$
in the diagram Fig.(\ref{one} e) \cite{lut}. Then the all diagrams
are readily simplified  with the following result for $\Gamma(\cos
\Theta)\equiv \Gamma_a+\Gamma_b+\Gamma_c+\Gamma_d+\Gamma_e$ in the
eigenvalue equation (\ref{lambda})
\begin{eqnarray}
&&\Gamma_a +\Gamma_b= {s\over{q_-^2 +s(q_-)}}, \cr
&&\Gamma_c+\Gamma_d ={4s^2\over{[q_-^2 +s(q_-)] q_-}}\int_0^{1} {k
dk \over{(k^2-q_-^2/4+4s)}} \times \cr && \left[\ln
{q_-/2+k\over{q_-/2-k}}-\ln
{k^2+q_-^2/4+4s+kq_-\over{k^2+q_-^2/4+4s-kq_-}} \right],   \cr
&&\Gamma_e = {s\over{4q_+}} [{1-q_+^2/4\over{1-q_+^2/4+4s}}\ln
{1+q_+/2\over{1-q_+/2}} -\cr
&&{1-q_+^2/4-4s\over{2(1-q_+^2/4+4s)}}
\ln{(1+q_+/2)^2+4s\over{(1-q_+/2)^2+4s}}],  \label{gamma2}
\end{eqnarray}
where $q_\pm^2=2(1\pm\cos \Theta)$.

\begin{figure}[tbp]
\begin{center}
\includegraphics[angle=-00,width=0.50\textwidth]{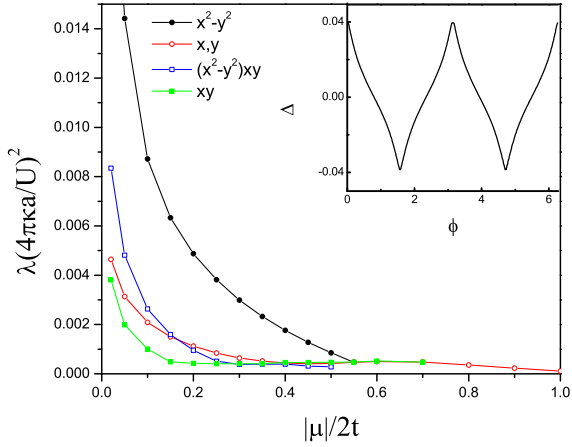}
\end{center}
\caption{(Color online) The absolute value of the coupling
constant for the 2D Hubbard model with the nearest-neighbor
hopping as the function of the chemical potential for four
unconventional symmetries of the order parameter. Inset shows the
angular dependence of the $d_{x^2-y^2}$ state similar to that
found in Ref. \cite{kivsca}.} \label{two}
\end{figure}

 Eqs.(\ref{gamma2},\ref{lambda}) allows us to find the symmetry (i.e. the angular momentum $l$) of the
order parameter with the most negative value of $\lambda$, shown
in Fig.(\ref{one}) as the function of the interaction strength. As
in the case of  independent screening we do not observe the p and
d-wave pairing in the whole region  of the Kohn-Luttinger
perturbation expansion and beyond (up to $s=3$)  in agreement with
Refs.\cite{lut,alegol}. There is a pairing in higher momentum
states, $l \geqslant 3$, but the corresponding eigenvalues  are
numerically so small ($\lambda_3\approx 0.0011$ for $s=3$), that
the corresponding T$_c \approx (E_F/k_B) \exp(-1/\lambda)$ is
virtually zero for any realistic Fermi energy $E_F$.

Finally, let us analyse the unconventional pairing in  a
two-dimensional electron gas on the square lattice  with a
tight-binding energy dispersion,
\begin{equation}
E_\textbf{p}=-2t [\cos(p_x a/\hbar)+\cos(p_y a/\hbar)]-\mu
\end{equation}
($a$ is the lattice constant). For the half-filled band the Fermi
level, $\mu$ is found at the van-Hove singularity (vHs) of the
density of states, $\mu=0$, so that one might expect a strong
enhancement of the unconventional T$_c$ near half-filling due to
vHs proximity \cite{kivsca}. The repulsion between electrons is
modeled by a strongly screened Coulomb potential with the Fourier
component $2\pi e^2/\Omega(q+\kappa)\approx 2\pi
e^2(1-q/\kappa)/\Omega\kappa$, where the inverse screening length
is taken as an independent large parameter, $\kappa a \gg 1$.

Similar to the 3D case,  the short screening length allows one to
neglect $q$ dependence of the potential  in the second-order
 diagrams slightly overestimating their contribution, while taking $q$ into account  in the first-order contribution.
 Then the diagrams b,c,d in Fig.(\ref {one}) cancel each other.
The remaining second-order  contribution (e),  proportional to the
static susceptibility of the noninteracting electrons,  can be
reduced to a one-fold integration, so that the two-particle vertex
contribution to the unconventional pairing is  expressed as
$K(\textbf{p},\textbf{p}^\prime) =(2\pi e^2/(\Omega a \kappa^2))
\Gamma(a\textbf{p}/ \hbar, a\textbf{p}^\prime/\hbar)$ with
 \begin{eqnarray}
&&\Gamma(\textbf{k},\textbf{k}^\prime)
=-|\textbf{k}-\textbf{k}^\prime| - \cr && {U\over {4\pi}}
\int^{k_m}_{-k_m}{dx\over{b}}\ln \left
|{(a_1-b)(a_2+b)\over{(a_1+b)(a_2-b)}}\right |, \label{2DK}
 \end{eqnarray}
 $a_{1,2}= \sin(x+q_y/2)\sin (q_y/2)\tan (\pm z/2 +
q_x/4)+\sin(q_x/2)$, $b=[\sin (q_x/2)^2-\sin (x+q_y/2)^2 \sin
(q_y/2)^2]^{1/2}$, $z=\cos^{-1}[-\cos(x)-\tilde{\mu})]$,
$\tilde{\mu}=\mu/2t$, $\textbf{q}=\textbf{k}+\textbf{k}^\prime$,
and $k_m=\cos^{-1}(-1-\tilde{\mu})$. Here one assumes that $|\sin
(q_x/2)|> |\sin (q_y/2)|$ and if otherwise, one should replace
$q_x\leftrightarrows q_y$.  No matter what the screening length
is, the relative contribution of the second-order diagram (e)
depends on a single dimensionless interaction parameter $U\equiv
e^2 /at \approx r_s$, which is supposed to be small in the
framework of the KL approach.

\begin{figure}[tbp]
\begin{center}
\includegraphics[angle=-00,width=0.50\textwidth]{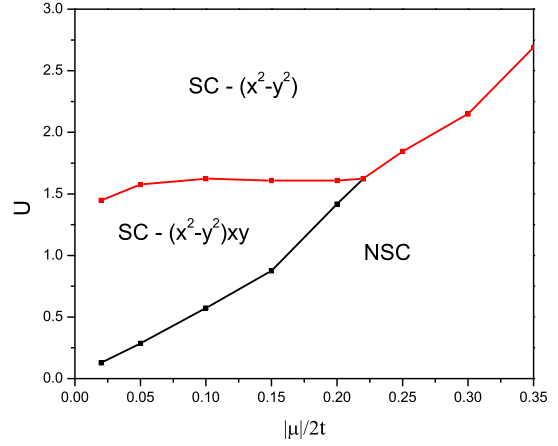}
\end{center}
\caption{(Color online)  $U-\mu$ phase diagram of  2D electrons on
the square lattice showing no d- or p-wave symmetry of the ground
state for weak repulsion $U$. There is a critical value of $U$ at
any doping below which the ground state is not superconducting
(NSC). } \label{three}
\end{figure}

Using Eq.(\ref{2DK}) we can solve the eigenvalue problem,
Eq.(\ref{bcs}), numerically as in Ref. \cite{kivsca} by
discretization of the Fermi surface and diagonalization of  the
kernel in the following integral equation,
\begin{equation}
\lambda \Delta(\textbf{k})={ U \over{4\pi  (\kappa
a)^2}}\int^{k_m}_{-k_m}dk_x^\prime
{\Gamma(\textbf{k},\textbf{k}^\prime)\Delta(\textbf{k}^\prime)+\Gamma(\textbf{k},-\textbf{k}^\prime)\Delta(-\textbf{k}^\prime)\over{|\sin
(k_y^\prime) \sin (k_y)|^{1/2}}}.
 \label{bcs2}
\end{equation}
Here vectors $\textbf{k}$ and $\textbf{k}^\prime$ are taken on the
Fermi surface, so that $k_y^\prime$ in the integral is defined via
$k_x^\prime$ using $\cos (\pi k_y^\prime) +\cos (\pi k_x^\prime)
=-\tilde{\mu}$.

\begin{figure}[tbp]
\begin{center}
\includegraphics[angle=-00,width=0.50\textwidth]{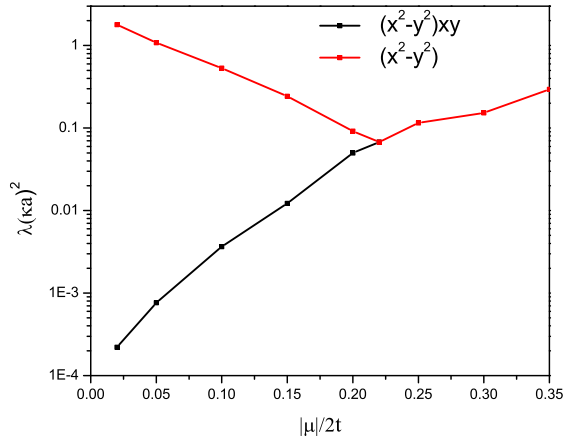}
\end{center}
\caption{(Color online) The absolute value of the coupling as the
function of the chemical potential for two unconventional order
parameters. } \label{four}
\end{figure}

With 800 discretization points we reproduce fairly well the
results of Ref.\cite{kivsca} for the 2D Hubbard model,
 if we drop the first term in $\Gamma(\textbf{k},\textbf{k}^\prime)$, Eq.(\ref{2DK})
 arising from the expansion of the bare Coulomb potential in powers of $1/\kappa$, Fig.(\ref{two}).
 As shown in the inset of Fig.(\ref{two}) the ground state is $B_{1g}$ spin singlet with the d-wave symmetry $x^2-y^2$ close to the half-filling  for any weak Hubbard repulsion.

 In fact, there is no reason to neglect the first-order q-correction  in the two-particle vertex Eq.(\ref{2DK}),
 since this correction is larger than the spin-fluctuation contribution at any screening, if the repulsion is weak.
 Using the correct vertex Eq.(\ref{2DK}) qualitatively changes the ground state, Fig.(\ref{three}).
 Contrary to Ref. \cite{bar,kivsca,kiv} neither p- nor d-wave  pairing are possible in the ground state at any filling until the effective interaction
 becomes so strong that the perturbation theory does not apply,  $U \gtrsim 1$, Fig.(\ref{three}), similar to the 3D case discussed above.
As in the continuum 3D Coulomb gas, Fig.(\ref{one}), there is
a pairing in higher momentum states (e.g. $A_{2g}$ with the
symmetry $(x^2-y^2)xy$),  but  as shown in Fig.(\ref{four}) the
corresponding eigenvalues  are numerically very small, and the
corresponding T$_c$ is about zero.

In cuprate superconductors and many other metallic compounds the
Coulomb interaction is rather strong $r_s \gg 1$, so that the
perturbative KL approach might have no direct relevance to these
materials. Different numerical techniques have been applied to
elucidate the ground state of the repulsive Hubbard model in the
intermediate to strong-coupling regime, $U > 1$ sometimes with
conflicting conclusions. In particular, recent studies by Aimi and
Imada \cite{imada}  using a sign-problem-free Gaussian-basis Monte
Carlo (GBMC) algorithm  showed that the simplest Hubbard model
with the nearest-neighbor hopping has no superconducting
condensation energy   at optimum doping. This striking result was
confirmed in the variational Monte Carlo (vMC) studies by
Baeriswyl \emph{et al.} \cite{bae}, who found, however, some
condensation energy away from the optimum doping and also adding
next-nearest neighbor hoppings. Importantly,  a similar vMC method
\cite{hardy}
 found that even a relatively weak finite-range electron-phonon
interaction with the BCS coupling constant $\lambda \approx 0.1$
induces a d-wave superconducting state in strongly correlated
metals with the condensation energy several times larger than can
be obtained with the Hubbard repulsion alone. Moreover, the
unconventional superconductivity has been shown to exist due to a
finite-range electron-phonon interaction \cite{hardy,alesym}
without the need for additional mechanisms such as spin
fluctuations.

In conclusion, we have shown that the p- and d-wave Cooper pairing
from the weak Coulomb repulsion  is not possible  between fermions
at any screening length and in any dimension.  Pairing in higher
momentum states  $(l \geqslant 3)$  has
 virtually zero T$_c$ for any realistic Fermi energy. The
unconventional pairing from the strong Coulomb repulsion is not
possible either since the corresponding condensation energy, if
any, is several times lower than the condensation energy caused by
the electron-phonon interaction.

We would like to thank Maxim Kagan, Viktor Khodel, and Dragan
Mihailovic for helpful discussions. The research was supported  by
the Royal Society.

\end{document}